\documentclass[11pt,a4paper]{article}

\usepackage[T1]{fontenc}
\usepackage[utf8]{inputenc}
\usepackage{lmodern}
\usepackage[a4paper,margin=2.5cm]{geometry}
\usepackage{microtype}
\usepackage{graphicx}
\usepackage{authblk}
\usepackage{titlesec}
\usepackage[numbers,round,sort&compress]{natbib}
\usepackage[hidelinks]{hyperref}
\usepackage[labelfont=bf,labelsep=period,font=small]{caption}

\titleformat{\section}{\large\bfseries}{}{0pt}{}
\titlespacing*{\section}{0pt}{0.6em}{0.3em}
\providecommand{\doi}[1]{\url{https://doi.org/#1}}

\newcommand{\paperfigure}[3]{%
  \begin{figure}[t]\centering
    \includegraphics[width=\linewidth]{#1}
    \caption{#3}\label{#2}
  \end{figure}}

\title{Grokipedia stretches the US political spectrum}
\author[a,b,*]{Philipp Koch}
\author[c]{Alexander Posztos}
\affil[a]{EcoAustria -- Institute for Economic Research, Vienna, Austria}
\affil[b]{Center for Collective Learning, IAST, Toulouse School of Economics, Universit\'e de Toulouse Capitole, Toulouse, France}
\affil[c]{TU Wien, Vienna, Austria}
\affil[*]{Corresponding author: \texttt{philipp.koch@ecoaustria.ac.at}}
\date{}

\begin{document}
\maketitle

\begin{abstract}
Language models now write reference works that people and other AI systems consult. Grokipedia, written by xAI’s model, presents itself as a less biased alternative to crowd-edited Wikipedia. We compare the biographies of 1,614 Democratic and Republican current and former politicians, from Andrew Jackson to members of the current Congress, on both encyclopedias in February and November 2026. For nine in ten Republicans, Grokipedia's biography reads as if written by a more conservative author than Wikipedia's. For Democrats, this holds only about half the time. We find the politicians’ ideology explains this asymmetry. For 1,022 politicians with Congressional roll call voting records, how conservative the author seems rises linearly with the politician's ideology and about twice as steeply on Grokipedia as on Wikipedia. Republicans in races where Elon Musk’s America PAC spent money show no extra bias. Rather than favoring a party, Grokipedia stretches the political spectrum, moving the biographies of the most liberal politicians further left and of the most conservative politicians further right.


\end{abstract}

\section{Introduction}

In April 2019, Mike DeWine, the Republican governor of Ohio, signed the ``Heartbeat Bill'', prohibiting abortion once cardiac activity can be detected in an embryo. Wikipedia reports the signing in a neutral tone. Grokipedia, the encyclopedia written by xAI’s language model, opens its version of the sentence with ``A longstanding pro-life advocate''. Asked whether the author supports conservatives, Political DEBATE~\cite{r1}, a classifier for political text, gives probabilities of 0.000005 for the Wikipedia sentence and 0.9999996 for the Grokipedia one (Fig.~\ref{fig:1}a). Is this only one example, or a pattern?

Media bias has been traced to what readers demand~\cite{r2} and what owners of media outlets choose~\cite{r3}. On Wikipedia, articles that are revised more often show less bias~\cite{r4}, which suggests that a crowd of editors corrects what single editors miss. Language models replace that crowd with a single author, and their political leaning can be shifted by fine-tuning~\cite{r5}. Unlike a private chatbot answer, however, an encyclopedia article is public and conveys legitimacy, so its author’s leaning may travel further.

Grokipedia launched in October 2025 with about 885,000 articles and had been cited by AI systems about 356,000 times by March 2026~\cite{r6}. It has not reviewed user edits since around April 2026~\cite{r6}. Early studies find it longer than Wikipedia, framed further to the right and less neutral on politicians~\cite{r7,r8,r9,r10}. Here, we ask what produces these differences: party favoritism, ideological amplification, or the politics of its owner, Elon Musk. To do so, we compare Wikipedia and Grokipedia biographies of 1,162 politicians in February and 1,614 politicians in November 2026, five days after the midterm elections. The sample is largely historical, as 959 of them held no office by then. We link 1,022 politicians (488 of them currently active) to the first dimension of DW-NOMINATE~\cite{r11,r12}, a left--right score from roll call voting records in Congress that is independent of either text.

\section{Results}

\textbf{Grokipedia shifts nine in ten Republican biographies right, but only half of Democratic ones}. We define lean as the mean probability that the author of a sentence supports conservatives minus that for liberals, and lean shift as Grokipedia’s lean minus Wikipedia’s (Materials and Methods). In February, Grokipedia shifts 89\% of the 572 Republican biographies to the right (Fig.~\ref{fig:1}b), by 0.126 on average (Fig.~\ref{fig:1}c, 95\% CI 0.117--0.135). DeWine is a typical case (lean 0.056 to 0.182, against 0.053 to 0.180 for the average Republican). Democrats barely move (0.001, CI $-$0.006 to 0.007), with 53\% of the 590 Democratic biographies moving right. For comparison, we compute the lean shift for all 98,013 biographies we could match: the shift averages only 0.002.

\paperfigure{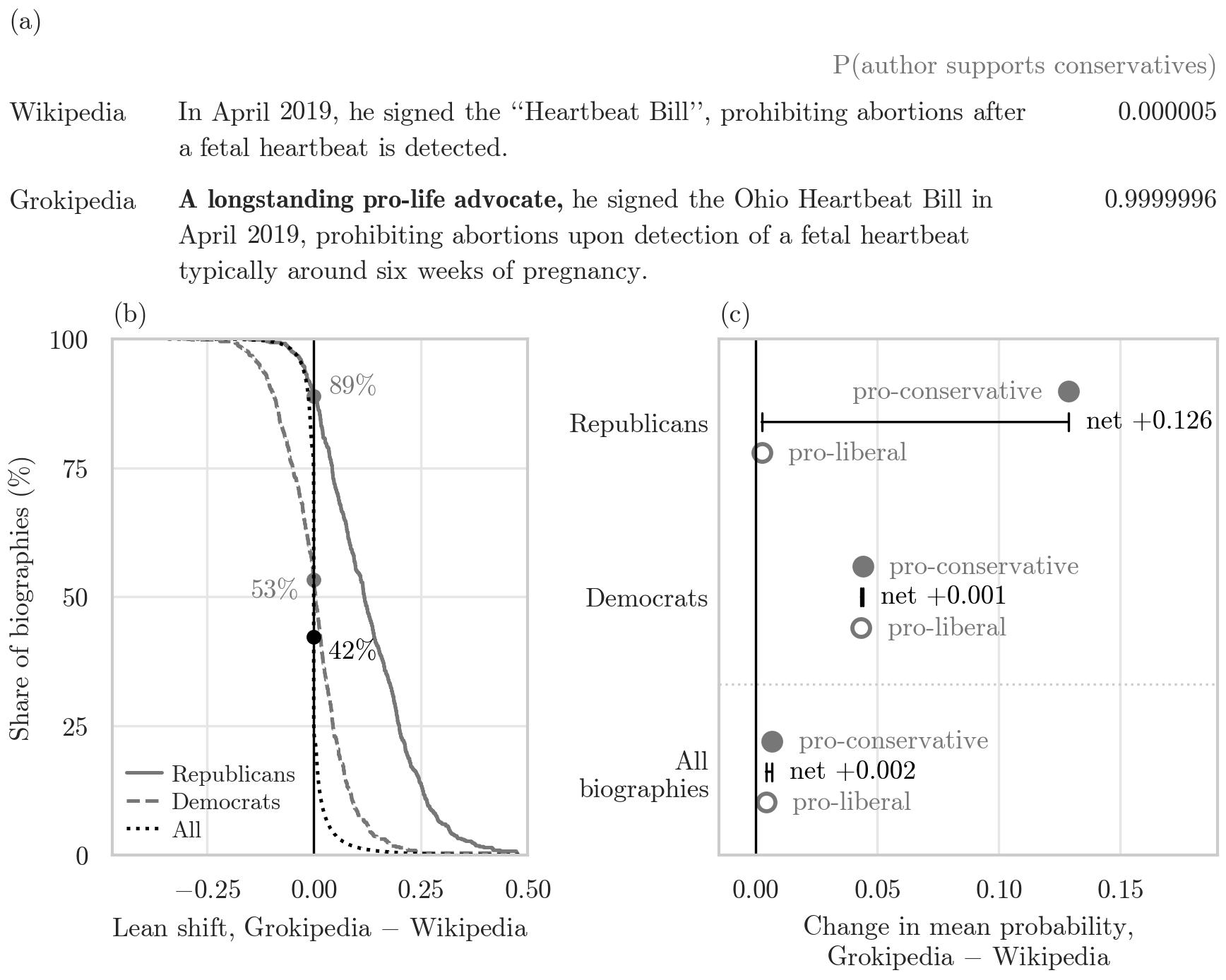}{fig:1}{Grokipedia moves Republican biographies right and leaves Democratic ones in place, on average. (a) The Mike DeWine sentence pair with the classifier probabilities that the author supports conservatives. (b, c) 1,162 politicians and all 98,013 biographies (dotted). (b) Share of biographies whose shift is at least x (complementary cumulative distribution); dots at x = 0 give the share moving right. (c) Change in the mean probability that the author supports conservatives (filled) or liberals (hollow). The bracket is the net shift.}

Decomposing the lean shift into components shows how this happens. In Republican biographies, Grokipedia raises the mean probability that the author supports conservatives but barely that for liberals (+0.129 against +0.002), which makes them more one-sided (Fig. 1c). In Democratic biographies, it raises both (+0.044 and +0.043, about 7 and 10 times the rise across all 98,013 biographies), so their average lean remains unchanged. A dictionary check without a language model agrees. The typical phrases of Republican speech in Congress~\cite{r13} become more frequent in Grokipedia’s Republican biographies (\textit{P} < 10\textsuperscript{$-$14}), but not in Democratic ones (\textit{P} = 0.52).

The pattern survives an apparent partial rewrite of Grokipedia in March 2026~\cite{r6}. In November, the shifts are 0.125 for the same Republicans and 0.000 for the same Democrats.

\textbf{Grokipedia stretches the political spectrum}. In November, Grokipedia’s lean rises with DW-NOMINATE at a slope of 0.437, against 0.210 on Wikipedia, a ratio of 2.08 (CI 1.99--2.17, Fig.~\ref{fig:2}a). Each step from liberal to conservative thus moves Grokipedia’s lean twice as far as Wikipedia’s. The shift combines an offset of 0.067 to the right with 0.227 per unit of ideology (Fig.~\ref{fig:2}b).

The ratio is robust: it is 2.02 among the 488 members of the current Congress. It lies between 1.94 and 2.33 within each party, in February alone, and in the third of pairs where Grokipedia’s article is closest to Wikipedia’s length.

\paperfigure{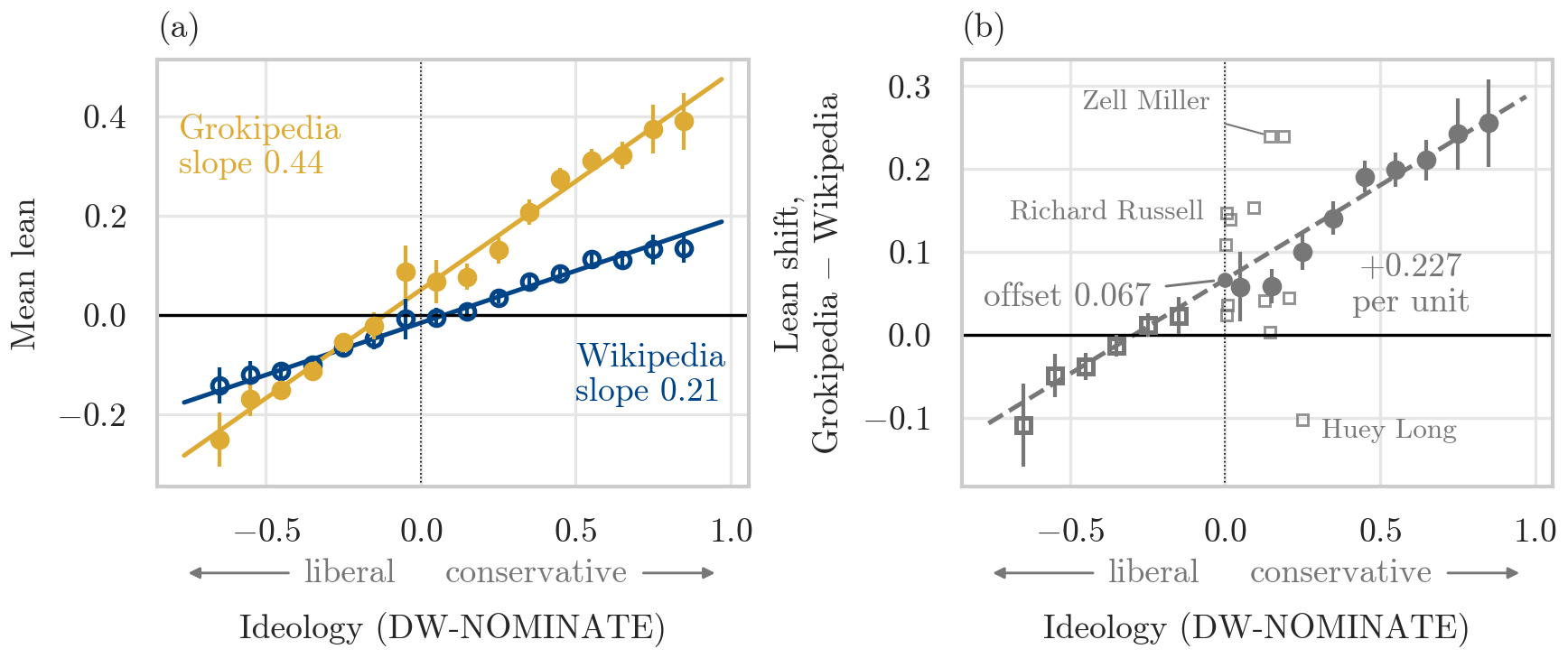}{fig:2}{Grokipedia's lean rises about twice as steeply with ideology as Wikipedia's. Lean is the mean probability that the author of a sentence supports conservatives minus that for liberals. Lean shift is Grokipedia's lean minus Wikipedia's. Sample: 1,022 politicians with a DW-NOMINATE score. Points are means of ideology bins of 0.1 with at least 10 politicians, bars 95\% CIs, lines least-squares fits to individuals. (a) Lean. (b) Shift for Democrats (squares) and Republicans (circles). Small squares: the 12 Democrats with a score above zero.}

\textbf{Races in which Musk’s PAC spent show no extra bias}. The 27 Republicans in races where Musk’s America PAC spent at least \$100,000 in 2024 or 2026 shift by 0.172, against 0.156 for 31 Republicans in comparably contested races without its money (difference 0.016, CI $-$0.039 to 0.072). The test rules out an extra shift larger than half of what other Republicans already receive (0.144). Grokipedia also did not edit these Republican articles more often during the campaign (22\% vs. 26\%).

\section{Discussion}

Grokipedia presents itself as a less biased encyclopedia. Relative to Wikipedia, however, Grokipedia stretches the political spectrum: it pushes liberals to the left and conservatives to the right. This may reflect a model that writes each politician in the voice of their supporters~\cite{r14}. We find no evidence that Grokipedia favors the politicians Musk supported.

Three limitations apply. First, the classifier cannot tell taking sides from describing positions. Wikipedia’s own slope of 0.210 shows that the classifier responds to described positions, while DeWine’s sentence points to framing. Only sentence-level annotation can separate the two. Second, DW-NOMINATE covers only members of Congress, so 592 of the 1,614 politicians have no score. Third, we cannot tell whether other language models would write biographies the same way.

Grokipedia thus gives its readers, and the AI systems that cite it~\cite{r6}, a more polarized picture of US politicians, past and present. We should therefore treat Grokipedia, and AI-written encyclopedias more generally, with care, because they may amplify ideological differences.

\section{Materials and Methods}

\textbf{Data. }The February snapshot, scraped on 11--12 February 2026, pairs the English Wikipedia and Grokipedia biographies of the people in the 2.0 release of Pantheon~\cite{r15}, which covers people with a presence in at least 15 language editions of Wikipedia. We identify 98,013 pairs among the two datasets. All serve as a reference in Fig.~\ref{fig:1}b. Of these, 1,162 are US politicians with exactly one of the two major parties among their Wikidata affiliations, which leaves out party switchers. The November snapshot (8 November 2026) rescrapes them and adds 452 current members of Congress, governors, cabinet members, mayors of large cities and the major-party nominees for Senate, House and governor in 2026 (which are not listed in Pantheon). We identify them through Wikipedia pages listing the respective politicians and results of the 2026 midterm elections.

\textbf{Political shift. }We split texts into sentences with spaCy and keep those with at least five words. Political DEBATE~\cite{r1}, a zero-shot entailment classifier pinned to a fixed model version, scores each sentence against the hypotheses ``The author of this text supports conservatives/liberals.'' As absolute probabilities are small for encyclopedic prose, we interpret only differences between the encyclopedias. The dictionary check scores stemmed bigrams against the partisan phrases of the 104th--114th Congresses~\cite{r13}.

\textbf{Ideology and spending. }We match politicians to Voteview~\cite{r12} via Wikidata bioguide identifiers and use the first dimension of DW-NOMINATE~\cite{r11}, averaged over each career. Classical measurement error in DW-NOMINATE attenuates both slopes equally and cancels in their ratio. America PAC reported more than \$270 million in independent expenditures in the 2024 and 2026 cycles (Federal Election Commission filings). We link them to candidates via Wikidata FEC identifiers and names. A race counts as Musk-backed with at least \$100,000 of America PAC spending and as contested with at least \$1 million of other outside spending.

\textbf{Data and code availability. } The replication package is available at \url{https://github.com/philmkoch/grokipedia-replication}. The article texts are available for research use only under the licenses that govern them (CC BY-SA 4.0 for Wikipedia, the xAI Community License for Grokipedia).

\section{Acknowledgments}

\textbf{Competing interests}. The authors declare no competing interests.

\textbf{Funding}. This research received no specific grant from external funding agencies.

\textbf{Author contributions}. Philipp Koch (Conceptualization, Validation, Writing -- Original Draft Preparation, Writing -- Review \& Editing, Visualization, Supervision), Alexander Posztos (Conceptualization, Methodology, Software, Formal Analysis, Data Curation, Writing -- Original Draft Preparation, Visualization).


\bibliographystyle{pnas-new}
\bibliography{refs}
\end{document}